\documentclass[aps,prl,floatfix,epsfig,twocolumn,showpacs,preprintnumbers]{revtex4}
\usepackage{amssymb}
\usepackage{graphicx}
\usepackage{amsmath}

\begin{document}

\title{Calculated Magnetic Exchange Interactions in High-Temperature
Superconductors}
\author{Xiangang Wan$^{1}$, Thomas A. Maier $^{2}$, Sergej Y. Savrasov$^{3}$}
\affiliation{$^{1}$National Laboratory of Solid State Microstructures and Department of
Physics, Nanjing University, Nanjing 210093, China}
\affiliation{$^{2}$1Center for Nanophase Materials Sciences and Computer Science and
Mathematics Division, Oak Ridge National Laboratory, Oak Ridge, Tennessee
37831}
\affiliation{$^{3}$Department of Physics, University of California, Davis, One Shields
Ave, Davis, CA 95616}
\date{\today}

\begin{abstract}
Using a first principles linear response approach, we study the magnetic
exchange interactions J for a series of superconducting cuprates. We
reproduce the observed spin-wave dispersions together with other
experimental trends, and show that different cuprates have similar J's
regardless their T$_{c}$. The nearest neighbor J is not sensitive to the
hole-doping, which agrees with recent experiments. For the undoped cuprates,
the second nearest neighbor J is ferromagnetic, but changes its sign with
hole-doping. We also find that, in contrast to the hopping integral, the
exchange interaction is not sensitive to the position of apical oxygen. To
see the effect of the long--range nature of the exchange on the
superconducting T$_{c}$, we study the dynamical spin susceptibility $\chi
(q,\omega )$\ within the t--J model using a dynamical cluster approximation.
\end{abstract}

\date{\today}
\maketitle

\section{INTRODUCTION}

After more than two decades of intensive studies of high--temperature
superconductors (HTSCs),\ it is now\ well accepted that magnetic exchange
interactions\ in the cuprates play a fundamental role. Despite a general
lack of consensus on the pairing mechanism in the cuprates, it is believed
to be of magnetic origin \cite{Pines,Monien,Anderson,SO5,SFtheory}. In
support of this scenario, a direct relationship $T_{c}\propto J$ between $%
T_{c}$ and the in-plane exchange interaction $J$ was extracted from magnetic
measurements for one family of the cuprates in Ref.~\cite{Kanigel}. Despite
vast efforts devoted to understanding the magnetic properties of the
cuprates \cite{Dai}-\cite{Mag}, there are still several important issues
which need to be clarified. Among them is the long range nature of the
exchange interactions $J$\ between the Cu-ions, and its effect on the
dynamical spin susceptibility $\chi (q,\omega )$ and the critical
temperature\ T$_{c}$. Studies of model Hamiltonians, such as the t--J model
show a strong dependence of the calculated properties on the value of $J$%
\cite{J is important,J-Tc}. In particular, numerical calculations suggest a
direct relationship\ between the magnitude of the magnetic exchange coupling
and the pair binding energy\cite{J-Tc}. Therefore,\ obtaining accurate
estimates for the exchange constants and studying their trends across
different HTSCs including their doping dependence and influence on $\chi
(q,\omega )$\ and T$_{c}$\ is an interesting problem which we address in the
present work.

Experiments provide estimates for the magnitude of the exchange interaction,
but despite\ many studies there is a spread in obtained values even for the
nearest--neighbor magnetic coupling $J_{1}\ $in the most studied compound La$%
_{2}$CuO$_{4}$ where it varies from 110 to 150 meV \cite{LaCuO-1}-\cite{SW
exp}. The latter result does not only depend on the experimental technique
being used but also on how the results are fitted, i.e., depending on
whether higher order exchange terms are included. Nevertheless, based on
recent\ high--resolution inelastic neutron scattering (INS) experiments,
there is an overall agreement that at least for the undoped La$_{2}$CuO$_{4}$%
\cite{SW exp}, a simple nearest--neighbor Heisenberg model is not sufficient
and the high-order magnetic exchange interaction is quite important.
However, all high--order terms have similar effects on the spin--wave
dispersion and its intensity dependence, therefore, even INS, the most
powerful technique for exploring magnetic excitations, cannot determine the
magnetic exchange coupling very accurately. Worse than La$_{2}$CuO$_{4}$,
for most other HTSC compounds, the difficulty in synthesizing large single
crystals limits the capability of performing accurate measurements. Thus,
the experimental\ information on the exchange interactions is very limited.

In addition to the experiment, the magnetic interactions of the undoped
cuprates have been studied in a number of theoretical works\cite{ring term
theory 1},\cite{Theory LCO-1},\cite{Theory LCO-2},\cite{Theory LCO-3}.
However, most of them use a cluster approximation and map the total energy
differences obtained by a first principle calculation to a model Hamiltonian
which prevents detailed studies of the long range nature of the $J$'s. It
is, for example, well known that the long--range hopping integrals are very
important for the electronic properties in the cuprates\cite{Shen} and even
the effective third nearest neighbor hopping t"\ has a considerable effect
on the interatomic exchange\cite{t-t'-t''-U}. Therefore it is desirable to
re-investigate the exchange interaction in cuprates by a first principle
calculation.

In contrast to the undoped cuprates, investigations of $J$'s in doped HTSCs
are scarce, regardless the superconductivity actually happens only after
introducing doping. Based on a model calculation, Si \textit{et al.},
suggested that $J_{1}$ will decrease rapidly with doping\cite{Qimiao Si}. On
the other hand, the commonly used t--J model uses the same $J$ for different
doping levels\cite{J is important}, therefore study the effect of doping on
the exchange interaction based on a first principle calculation is quite
interesting. Unfortunately, \textit{ab initio} techniques for extracting $J$%
's (like the cluster approximations mentioned above) are not efficient for
the doped case, and there are no extensive calculations of this type
reported in the literature. In this work, we use a recently developed linear
response approach \cite{Wan J}, and perform detailed studies of exchange
interactions for both parent and doped HTSCs. Beyond this, we\ also
investigate the effect of the long--range J's\ on the spin susceptibility $%
\chi (q,\omega )$ and the superconducting transition temperature T$_{c}$
within the framework of the dynamical cluster approximation (DCA) for the
t--J model.

\section{EXCHANGE INTERACTION\ AND\ SPIN\ WAVES}

\subsection{Method}

We perform our electronic structure calculations based on density functional
theory (DFT)\ within the full potential linearized-muffin-tin-orbital (LMTO)
method\cite{Savrasov-1996}. To take into account the effect of the on--site
electron--electron interaction we supplement the local density approximation
(LDA) to DFT by adding a correction due to Hubbard U using so--called LDA+U
approach\cite{LDA+U}, with the parameters U$=$10 eV and J$=$1.20 eV for the
Cu $d$ orbitals as deduced from the constrained LDA calculation\cite{U J}.
Experimental lattice parameters have been used for all materials.

With the electronic structure information, one can evaluate the magnetic
exchange parameters $J$\ of a Heisenberg model $H=\sum_{ij}J_{ij}S_{i}\cdot
S_{j}$ based on a magnetic force theorem\cite{force theorem} which assumes a
rigid rotation of atomic spin. In this formalism, the interatomic exchange
constant $J$ is given as a second derivative of the total energy difference
induced by the rotation of moments at sites $R+\tau $ and $R^{^{\prime
}}+\tau ^{^{\prime }}$ \cite{Wan J}: 
\begin{eqnarray}
J_{\tau R\tau ^{^{\prime }}R^{^{\prime }}}^{\alpha \beta } &=&\sum_{\mathbf{%
\text{q}}}\sum_{\mathbf{\text{k }}jj^{^{\prime }}}\frac{f_{\mathbf{\text{k}}%
j}-f_{\mathbf{\text{k}+q}j^{^{\prime }}}}{\epsilon _{\mathbf{\text{k}}%
j}-\epsilon _{\mathbf{\text{k}+q}j^{^{\prime }}}}\langle \psi _{\mathbf{%
\text{k}}j}|[\sigma \times \mathbf{B}_{\tau }]_{\alpha }|\psi _{\mathbf{%
\text{k}+q}j^{^{\prime }}}\rangle  \notag \\
&\times &\langle \psi _{\mathbf{\text{k}+q}j^{^{\prime }}}|[\sigma \times 
\mathbf{B}_{\tau ^{^{\prime }}}]_{\beta }|\psi _{\mathbf{\text{k}}j}\rangle
e^{i\mathbf{\text{q}}\cdot (\mathbf{\text{R}}-\mathbf{\text{R}^{^{\prime }}}%
)},  \label{J}
\end{eqnarray}%
where $f$, $\sigma $, and \textbf{B} are the Fermi function, Pauli matrix,
and the effective magnetic field in the calculation given by the difference
in the electronic self-energies for spin-up and spin-down electrons,
respectively. $\psi $\ and\ $\epsilon $\ are the eigenstate and eigenvalue
from the LDA+U calculation. This technique has been used successfully for
evaluating magnetic interactions in a series of transition metal oxides\cite%
{Wan J}.

\begin{figure}[tbp]
\includegraphics[width=3.5in]{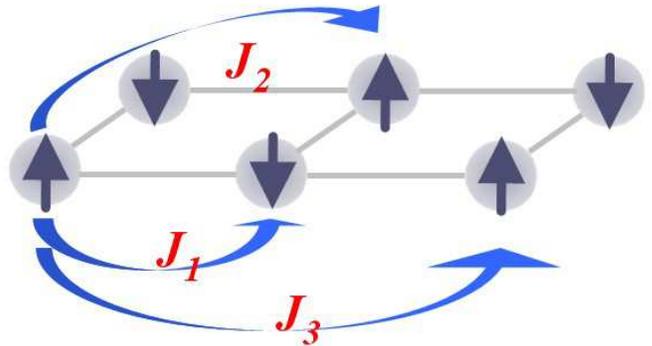}
\caption{Definitions of nearest neighbor, $J_{1}$, next nearest neighbor, $%
J_{2},$ and third nearest neighbor, $J_{3},$ exchange interactions for Cu
spins, which are the parameters of a Heisenberg model $H=\sum_{\langle
ij\rangle }J_{ij}S_{i}S_{j}$.}
\label{fig1}
\end{figure}

\subsection{Parent Compounds}

We now discuss our results for La$_{2}$CuO$_{4}$ and find that the exchange
constants $J_{1}$, $J_{2}$ and $J_{3}$ within the CuO$_{2}$ plane (see Fig.
1) decrease rapidly with increasing distance between two Cu ions. After
taking into account the effect of quantum renormalization\cite%
{Renormalization}, the measurement of two--magnon Raman scattering gives $%
J_{1}$=116 meV\cite{LaCuO-1}, which is slightly larger than our numerical $%
J_{1}$ (109 meV) as shown in Table I. Early neutron scattering experiments
give a larger value of $J_{1}\sim $130 meV\cite{LaCuO-2}\cite{LCO-3}, which
may be partially due to the use of only a nearest--neighbor Heisenberg model
to fit their spin-wave velocities. Our $J_{1}$ agrees very well with one
cluster calculation (105 meV)\cite{Theory LCO-1} but is smaller than the
result of other calculations ($\sim $140 meV)\cite{Theory LCO-2,ring term
theory 1}. Turning to the discussion of the next nearest neighbor coupling,
our calculated $J_{2}$ is slightly larger than the one deduced from neutron
scattering\cite{LaCuO-2}. In contrast to the early cluster calculations\cite%
{Theory LCO-3}, our $J_{2}$ is ferromagnetic (FM) and thus enhances the
antiferromagnetic correlations.

Since well--defined spin--wave excitations throughout the Brillouin Zone
have been observed by the INS\cite{SW exp}, it is interesting to perform the
comparisons with our calculated spin--wave dispersions. For cuprates the
quantum fluctuations may\ be large due to the smallness of the spin S=$\frac{%
1}{2}$ and the low dimensionality D=2. On the basis of the
Holstein--Primakoff transformation using 1/S expansion, it has been found
that a renormalization factor is necessary for the spin--wave excitation
energy\cite{Renormalization} in order to compare it with the result of
standard linear spin--wave theory\cite{Linear Spin-wave}. With this
correction, the spin--wave dispersion can be expressed as: 
\begin{equation*}
E_{q}=2Z_{c}\sqrt{A_{q}^{2}-B_{q}^{2}}
\end{equation*}%
where $A_{q}=J_{1}-J_{2}[1-\cos (2\pi q_{x})\cos (2\pi q_{y})]-J_{3}[1-\frac{%
1}{2}(\cos (4\pi q_{x})+\cos (4\pi q_{y}))],$ $B_{q}=\frac{1}{2}J_{1}[\cos
(2\pi q_{x})+\cos (2\pi q_{y})],$ and $Z_{c}$ is the renormalization factor,
respectively. Here based on the above formula with the quantum
renormalization factor $Z_{c}$=1.18\cite{Renormalization}, and using the
obtained numerical $J_{1}$, $J_{2}$, and $J_{3}$, we calculate the
spin--wave dispersion for La$_{2}$CuO$_{4}$, and display the result in Fig.2
by the solid line. For comparison, in Fig. 2 we also show the INS data by
symbols\cite{SW exp}. Our results agree well with the experiments near the
zone center. A small discrepancy exists around the zone boundary which may
be due to the\ four--particle cyclic exchange interaction $J_{r},$ which
recently attracted much attention\cite{Ring-term exp,ring term theory 1}.

\begin{figure}[tbp]
\includegraphics[height=2.5in]{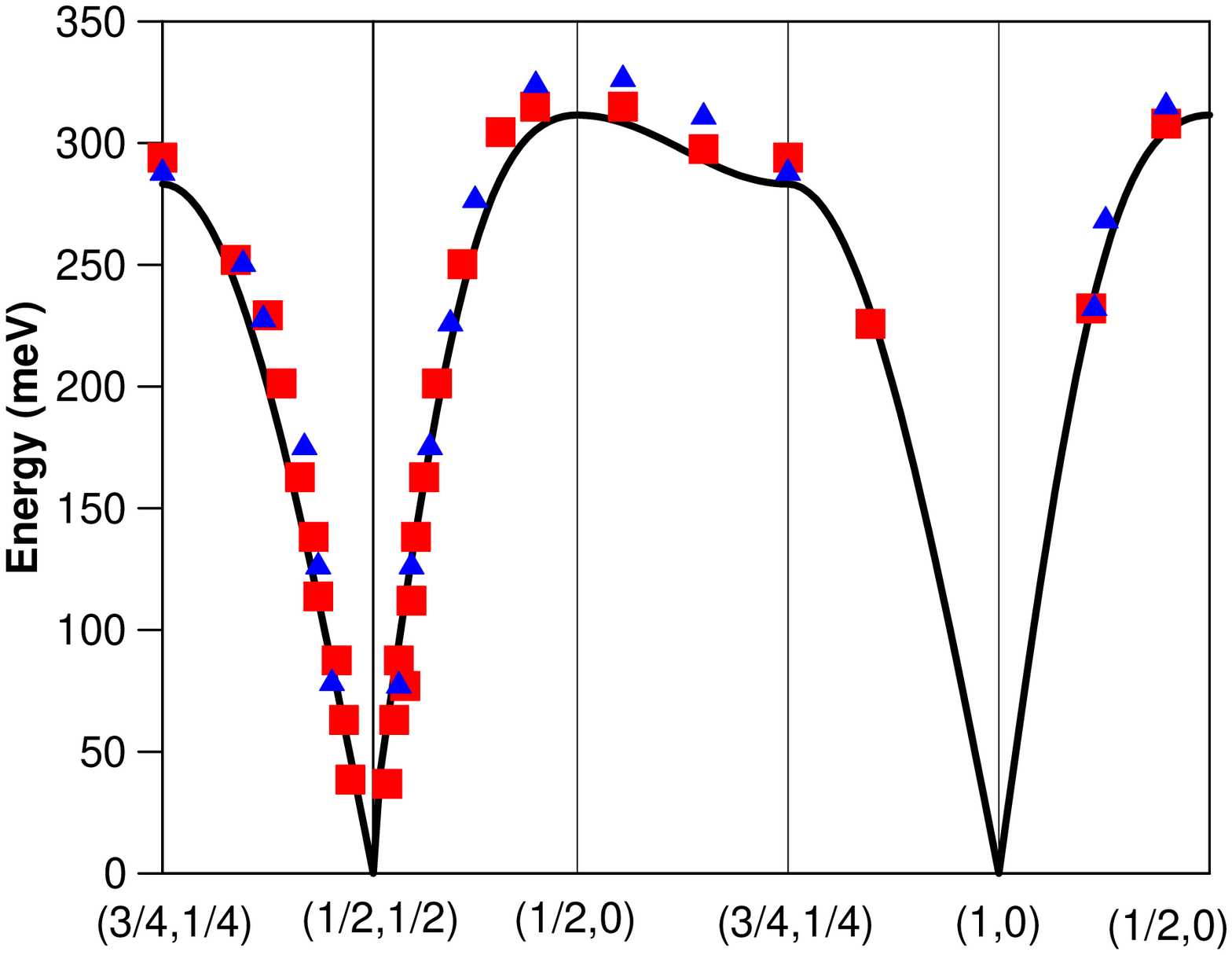}
\caption{Comparison between calculated (solid lines) and experimental
(symbols) spin-wave dispersions for La$_{2}$CuO$_{4}$. The triangles and
squares are the experimential results at T=10 K and 295 K, respectively 
\protect\cite{SW exp}.}
\label{fig2}
\end{figure}

We now turn to our predictions of exchange interactions for other HTSC
materials where the experimental values of $J$ are limited. The experiment%
\cite{Sr2CuOCl2}\cite{Keimer} shows the $J_{1}$ of Sr$_{2}$CuO$_{2}$Cl$_{2}$
is about 10 meV smaller than that of La$_{2}$CuO$_{4}$. This is reproduced
by our numerical calculation as shown in Table I. The $J_{1}$ of YBa$_{2}$Cu$%
_{3}$O$_{6.15}$ has been measured by neutron scattering, and the obtained
value is about 120 meV\cite{YBCO-1}\cite{YBCO-2}. After considering the
quantum renormalization effect\cite{Renormalization}, the experimental $%
J_{1} $ will be reduced to 100 meV, which is very close to our numerical
result 93 meV. It is interesting to note that our theoretical study
reproduces the experimental trend\ across the materials studied (La$_{2}$CuO$%
_{4}$ has largest $J_{1}$, Sr$_{2}$CuO$_{2}$Cl$_{2}$ is intermediate, and $%
J_{1}$ in YBa$_{2}$Cu$_{3}$O$_{6}$ is smallest\cite{Keimer}$.$) As it is
seen from Table I,\ all parent HTSCs have almost the same $J_{1}$ (around
110 meV) regardless of their number of CuO$_{2}$ layers and their different T%
$_{c}$'s. $J_{2}$ is also similar and shows FM\ behavior while $J_{3}$ is
AFM like. Using the quantum renormalization factor\cite{Renormalization} and
linear spin--wave theory, we also calculate the spin--wave dispersion for
all other compounds and show the results for HgBa$_{2}$CaCu$_{2}$O$_{6}$, Sr$%
_{2}$CuO$_{2}$Cl$_{2}$ and YBa$_{2}$Cu$_{3}$O$_{6}$ in Fig.3. Since
different compound have similar exchange interactions the shape of the
spin--wave curve is quite similar while YBa$_{2}$Cu$_{3}$O$_{6}$\ has
smaller spin--wave excitation.

\begin{figure}[tbp]
\vskip -2.5cm \includegraphics[height=4.5in]{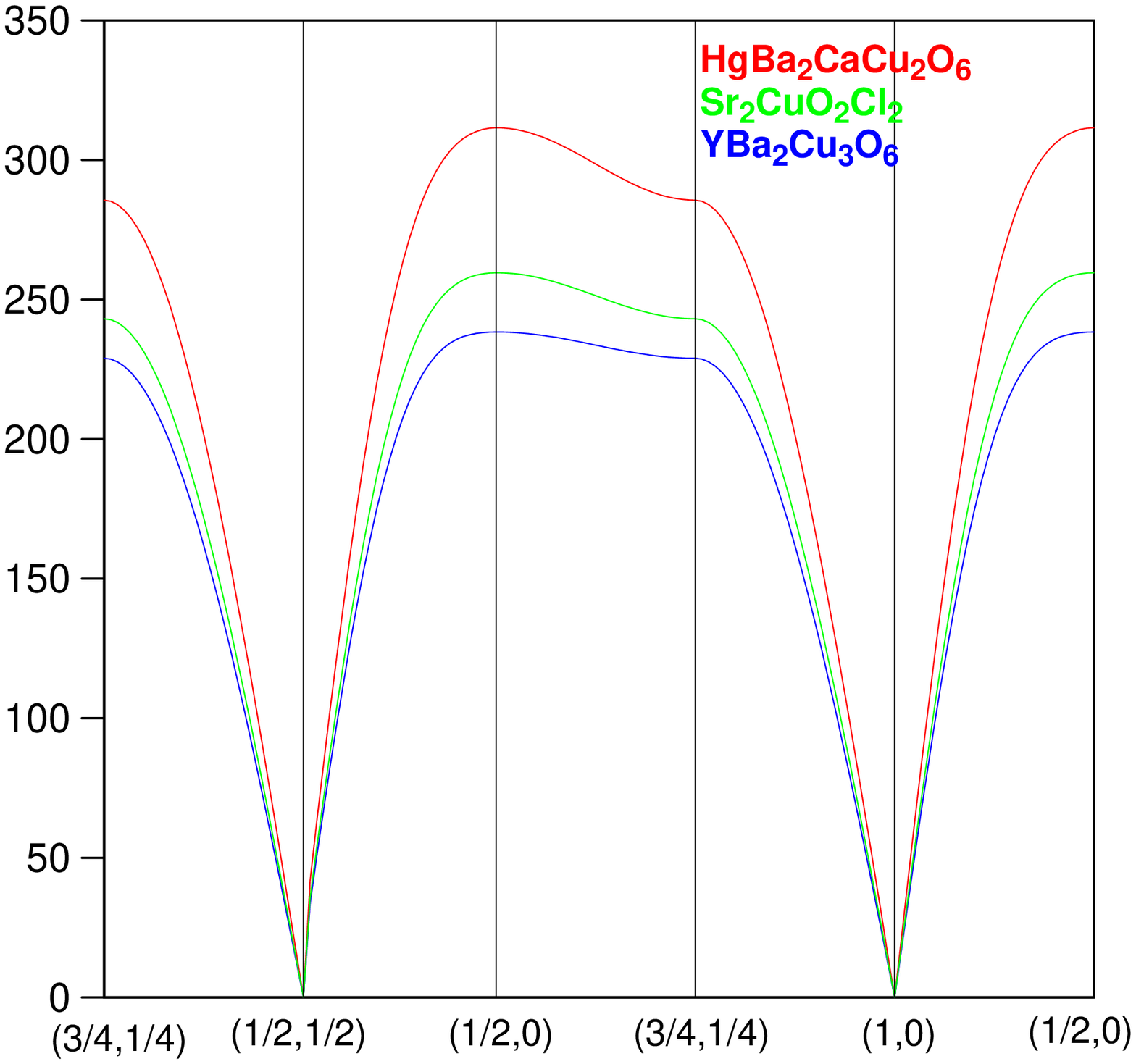} \vskip -2.5cm
\caption{Calculated spin--wave dispersions for HgBa$_{2}$CaCu$_{2}$O$_{6}$,
Sr$_{2}$CuO$_{2}$Cl$_{2}$ and YBa$_{2}$Cu$_{3}$O$_{6}$.}
\label{fig3}
\end{figure}

To further check the possible relationship between $J$\ and T$_{c}$, we also
study a high--pressure phase of HgBa$_{2}$Ca$_{2}$Cu$_{3}$O$_{8}$, for which
T$_{c}$ increases considerably\cite{P-Tc}. Applying pressure reduces the
lattice parameter, and, as a result, increases the magnitude of $J$, but
this enhancement is too weak to relate it with rising T$_{c}$. For example,
we find the values of $J_{1}$ are 114.3 and 116.4 meV at 3 and 8 GPa,
respectively, which is very close to the value at ambient pressure as shown
in Table I.

\subsection{The effect of apical oxygen}

It is believed that the apical oxygen has a dramatic effect on T$_{c}$\cite%
{Apical O 1,Apical O 2,Andersen PRL,W.Ku}. However, there is a debate on
whether T$_{c}$ is positively correlated with the magnitude of the effective
next nearest neighbor hopping $t^{\prime }$ \cite{Andersen PRL}, or if an
intersite super-repulsion term is important \cite{W.Ku}. We therefore study
the effect of the position of the apical oxygen on the exchange interaction.
The apical oxygen in La$_{2}$CuO$_{4}$ is located at the (0,0,z$_{o}$) site.
By adjusting the internal atomic coordinate z$_{o}$\ we perform the
calculation for different distances d$_{A}$ between the apical oxygen and Cu.

Our results are shown in Table II. In contrast to the hopping integral,
which is quite sensitive to the position of the apical oxygen\cite{Andersen
PRL,W.Ku}, the magnitude of the exchange interaction in the CuO$_{2}$ plane
only slightly increases with d$_{A}$\ as shown in Table II. In the Hg-based
cuprates with d$_{A}\sim 2.7$ \AA \thinspace\ $J$ is found smaller than the
one in La$_{2}$CuO$_{4}$\ with the same d$_{A}$, so it is likely that the
exchange interaction is more sensitive to the detail of the electronic
structure, rather than just to the distance d$_{A}$.

\begin{table}[tbp]
\caption{Experimental $T_{c}$ (K), and calculated exchange interactions
(meV)\ for parent HTSC materials. $J_{1}$, $J_{2}$, and $J_{3}$ are the
nearest--neighbor, second nearest--neighbor and the third nearest neighbor
exchange interactions as shown in Fig.1. N$_{layer}$ is the number of CuO$%
_{2}$ layer, T$_{c}$ is critical temperature.}%
\begin{tabular}{llllll}
\hline
& N$_{layers}$ & $T_{c}$ & $J_{1}$ & $J_{2}$ & $J_{3}$ \\ \hline
CaCuO$_{2}$ & 1 & -- & 110.0 & -10.1 & 3.8 \\ 
Tl$_{2}$Ba$_{2}$CuO$_{6}$ & 1 & 97 & 109.1 & -10.9 & 4.0 \\ 
HgBa$_{2}$CuO$_{4}$ & 1 & 94 & 108.9 & -11.1 & 3.3 \\ 
La$_{2}$CuO$_{4}$ & 1 & 42 & 108.8 & -12.0 & -0.2 \\ 
Sr$_{2}$CuO$_{2}$Cl$_{2}$ & 1 & 28 & 99.2 & -8.2 & 1.6 \\ 
HgBa$_{2}$CaCu$_{2}$O$_{6}$ & 2 & 128 & 110.4 & -11.9 & 2.9 \\ 
Tl$_{2}$Ba$_{2}$Cu$_{2}$O$_{8}$ & 2 & 125 & 108.7 & -10.7 & 2.5 \\ 
YBa$_{2}$Cu$_{3}$O$_{6}$ & 2 & 90 & 93.0 & -4.7 & 2.4 \\ 
HgBa$_{2}$Ca$_{2}$Cu$_{3}$O$_{8}$ & 3 & 135 & 109.9 & -10.1 & 2.8 \\ \hline
\end{tabular}%
\end{table}

\subsection{The effect of doping}

It may not be surprising that the $J$'s of the parent HTSCs do not directly
correlate with their T$_{c}$'s which characterize the corresponding doped
materials.\ We therefore study the effect of hole doping for La$_{2}$CuO$%
_{4} $ using the virtual--crystal approximation (VCA), which has been used
successfully for the phonon properties of La$_{2-x}$Ba$_{x}$CuO$_{4}$\cite%
{VCA}. For the doped case, our scheme is rough, but nevertheless does
include major ingredients of the system, such as, superexchange, double
exchange and RKKY exchange interactions \cite{Qimiao Si}. Naively, one may
think the hole induced by doping can hop through the Cu ion, and result in a
ferromagnetic like double--exchange interaction, which would consequently
suppress $J_{1}$. But if the doping level is not high, this effect is not
large as seen in Table III. Since the spin--wave velocity is mainly
controlled by $J_{1}$, our results agree with the recent INS experiments,
which showed clearly that for La$_{2-x}$Sr$_{x}$CuO$_{4}$ the spin--wave
velocity is doping insensitive \cite{hole-doped J}. Hole doping enhances $%
J_{3}$ slightly as seen in Table III. Different from $J_{1}$ and $J_{3}$,
doping has a large affect on $J_{2}$, which changes from FM--like to
AFM--like resulting in considerable spin fluctuation.

\begin{table}[tbp]
\caption{The calculated exchange interaction in La$_{2}$CuO$_{4}$, with
different d$_{A}$, where d$_{A}$ is the distance between apical oxygen and
Cu atom. d$_{A}$ is in \AA\ and $J$ is in meV.}%
\begin{tabular}{lllllll}
\hline
d$_{A}$ & $J_{1}$ & $J_{2}$ & $J_{3}$ &  &  &  \\ \hline
2.5 & 111.1 & -12.4 & -0.4 &  &  &  \\ 
2.6 & 112.9 & -13.1 & 0.1 &  &  &  \\ 
2.7 & 114.2 & -13.8 & 1.2 &  &  &  \\ 
2.8 & 116.0 & -14.6 & 2.1 &  &  &  \\ \hline
\end{tabular}%
\end{table}

\begin{table}[tbp]
\caption{Doping effect on exchange interactions in La$_{2}$CuO$_{4}$, where $%
x$ is the hole--doping concentration. $J$ is in meV.}%
\begin{tabular}{llll}
\hline
x & $J_{1}$ & $J_{2}$ & $J_{3}$ \\ \hline
0.0 & 108.8 & -12.0 & -0.2 \\ 
0.1 & 110.9 & -7.9 & -3.0 \\ 
0.2 & 117.8 & -0.3 & -3.1 \\ 
0.3 & 124.6 & 6.4 & -3.8 \\ \hline
\end{tabular}%
\end{table}


\section{Effect on $T_c$}

Now, we aim to address the following questions: (1) Can the variation in
exchange parameters for different materials explain the difference in $T_c$?
(2) What is the effect of varying exchange parameters on the dynamic spin
fluctuation spectrum? To answer these questions, we use a dynamic cluster
approximation (DCA) \cite{hettler:dca,ref:8} to calculate the properties of
a t-J model. A similar study was performed in Ref.~\cite{Kent}. There, a
combined DFT-LDA and DCA-QMC approach was used to study the parameter
dependence of $T_c$ in a three-band Hubbard model. These calculations showed
that $T_c$ is a very strong function of the hopping parameters, and very
sensitive to even small variations in the long-range hopping integrals.

Here, we want to study the dependence of $T_c$ on the exchange parameters $J$%
. We therefore consider a two-dimensional t--J model 
\begin{equation}
H=-t\sum_{\langle ij\rangle ,s}(\tilde{c}_{is}^{\dagger }\tilde{c}_{js}^{%
\phantom\dagger }+\tilde{c}_{is}^{\dagger }\tilde{c}_{js}^{\phantom\dagger
})+\sum_{ij}J_{ij}\mathbf{S}_{i}\mathbf{S_{j}}\,,  \label{eq:tJmodel}
\end{equation}%
where $\mathbf{S_{i}}=\tilde{c}_{is}^{\dagger }\mathbf{\sigma }_{ss^{\prime
}}\tilde{c}_{is^{\prime }}^{\phantom\dagger }$ and $\tilde{c}_{is}^{\dagger
} $ is a projected fermion operator defined as $c_{is}^{\dagger }(1-n_{i-s})$%
.

The general idea of the DCA is to map the bulk lattice problem onto an
effective periodic cluster embedded in a self--consistent dynamic host
designed to represent the remaining degrees of freedom. Correlations within
the cluster are treated explicitly, while those beyond the cluster size are
treated on the mean--field level. The hybridization of the cluster to the
host accounts for fluctuations arising from the coupling between the cluster
and the rest of the system.

The mean-field nature of the approach allows us to study transitions to
symmetry broken phases such as the superconducting state even in small
clusters. For example, using a dynamic cluster quantum Monte Carlo
approximation for a small four--site 2$\times $2 cluster, the properties of
a 2D Hubbard model were calculated in Ref.~\cite{jarrell_epl}. The obtained
phase-diagram is remarkably similar to the universal cuprate phase diagram,
exhibiting antiferromagnetic and $d$-wave superconducting phases as well as
pseudogap behavior. As discussed in Ref.~\cite{ref:8}, a four--site cluster
DCA calculation provides a mean-field result for the transition to a
superconducting state with a $d_{x^2-y^2}$-wave order parameter. DCA
calculations for larger cluster sizes show a reduction in the
superconducting $T_c$ due to the inclusion of pair-field phase fluctuations,
but the qualitative aspects of the results including the pairing mechanism
are similar to the four--site cluster results \cite{maier:pairmech}.

Here, we have used a non--crossing approximation \cite%
{maier:edca,maier:dcanca,haule07} (NCA) to determine the spin susceptibility 
$\chi (q,\omega )$ and $T_{c}$ for a four--site 2$\times $2 cluster. We
perform the simulations in the superconducting state by allowing for a
finite anomalous Green's function \cite{maier:dcanca}. With increasing
temperature, $T_c$ is determined by the temperature where the the anomalous
Green's function vanishes. Similar calculations for fixed near-neighbor
exchange integral $J=0.3t$ were performed in Ref.~\cite{haule07}. Many
results of this study were shown to be reminiscent of experiments in the
cuprates. Here, we focus on the effect of varying longer-ranged exchange
parameters on the spin-susceptibility and the superconducting $T_c$.

\begin{figure}[tbp]
\includegraphics[height=2.5in]{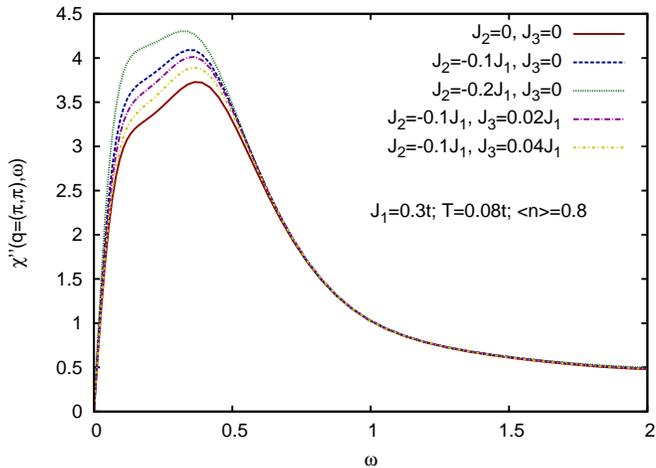}
\caption{Spin susceptibility $\protect\chi^{\prime \prime }(q,\protect\omega%
) $ for $q=(\protect\pi,\protect\pi)$ versus $\protect\omega$ for different
exchange interactions $J_2$ and $J_3$. Here, $J_1=0.3t$ is fixed and a
filling $\langle n\rangle = 0.8$ and temperature $T=0.08t$ have been used.}
\label{fig:4}
\end{figure}

Results showing the imaginary part of the spin susceptibility, $\chi
^{\prime \prime }(q,\omega )$ versus $\omega $ for $q=(\pi ,\pi )$,
calculated at a temperature $T=0.08t$ and fixed $J_{1}=0.3t$, are plotted in
Fig.~\ref{fig:4} for various values of the next--nearest--neighbor and
third--nearest neighbor exchange interactions $J_{2}$ and $J_{3}$. As one
can see, a ferromagnetic $J_{2}<0$ and an antiferromagnetic $J_{3}>0$
enhance the spectral weight in $\chi ^{\prime \prime }(q,\omega )$ at the
antiferromagnetic wave vector $q=(\pi ,\pi )$ at low frequencies.

Previous DCA quantum Monte Carlo and NCA simulations have addressed the
question of the pairing mechanism in the Hubbard and t--J models \cite%
{maier:pairmech}. The results of a recent DCA/NCA and Lanczos study of the
superconducting gap function were found to be consistent with a simple
phenomenological form for the $d$--wave pairing interaction \cite%
{maier:pairglue} 
\begin{eqnarray}
V_{d}(k,\omega ,k^{\prime },\omega ^{\prime }) &=&  \label{eq:pairint} \\
&&\hspace{-2.9cm}\frac{3}{2}\bar{U}^{2}\chi (k-k^{\prime },\omega -\omega
^{\prime })-\bar{J}(\cos k_{x}-\cos k_{y})(\cos k_{x}^{\prime }-\cos
k_{y}^{\prime })\,.  \notag
\end{eqnarray}%
Here, $\bar{U}$ and $\bar{J}$ are effective coupling constants of the
retarded spin fluctuation part and the non--retarded exchange contribution,
respectively. It was shown that the dominant contribution to the pairing
interaction $V_{d}$ comes from the spin fluctuations \cite{maier:pairglue}.

\begin{figure}[tbp]
\includegraphics[width=3.5in]{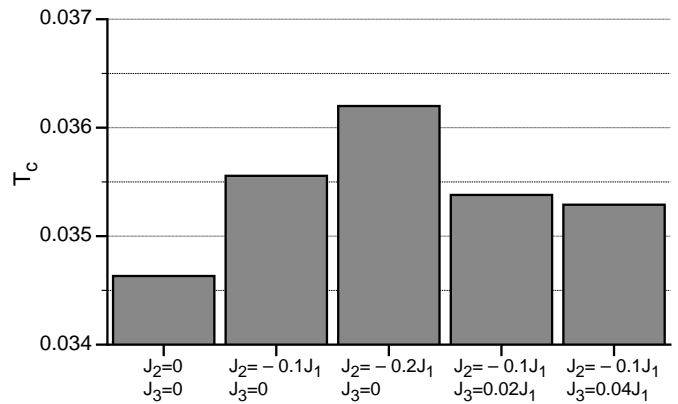} \centering
\caption{Superconducting transition temperature for different exchange
interactions $J_2$ and $J_3$. Here, we have used the same values for $J_1$, $%
J_2$ and $J_3$ and the filling as in Fig.~\protect\ref{fig:4}}
\label{fig:5}
\end{figure}

From this one would expect that variations in the spectral weight in $\chi
^{\prime \prime }(q,\omega )$ will directly affect the strength of $V_{d}$
and thus $T_{c}$. Fig.~\ref{fig:5} shows the DCA/NCA results for $T_{c}$ for
different values of the next--nearest--neighbor and third--nearest--neighbor
exchange integrals $J_{2}$ and $J_{3}$. Here, we have fixed the
nearest--neighbor exchange $J_{1}=0.3t$ and the filling $\langle n\rangle
=0.8$. Consistent with Eq.~(\ref{eq:pairint}), one finds that the magnitude
of $T_{c}$ tracks the magnitude of the spectral weight in $\chi ^{\prime
\prime }(q,\omega )$ for $q=(\pi ,\pi )$. E.g., the highest $T_{c}$ is
obtained for $J_{2}=-0.2J_{1}$, $J_{3}=0$, i.e. when the spectral weight in $%
\chi ^{\prime \prime }(q,\omega )$ is maximal for $q=(\pi ,\pi )$.

We emphasize, however, that $T_{c}$ is rather insensitive to changes in the
long--range exchange parameters $J_{2}$ and $J_{3}$. E.g. a change in $J_{2}$
from $J_{2}=0$ to $J_{2}=-0.2J_{1}$ only induces a $\approx 5\%$ increase in 
$T_{c}$ and the effects of $J_{3}$ on $T_{c}$ are almost negligible. Note,
however, that the third--nearest neighbor exchange $J_{3}$ is not directly
represented on a 4--site 2$\times $2 cluster. In this approximation, the
spins on the cluster only know about $J_{3}$ through their effective
coupling to the spin--fluctuation host. Therefore, it is possible that
larger variations in $T_{c}$ would be found in larger cluster simulations.
>From the present results, however, we conclude that the differences in the
long--ranged exchange terms cannot explain the differences in $T_{c}$
between different cuprates.

\section{CONCLUSION}

In summary, we have calculated the magnetic exchange interactions for a
series of HTSC cuprates using the LDA+U based linear response approach. Our
calculated spin wave dispersions were found in good agreement with
high-resolution\ inelastic neutron scattering experiments. Our simulations
show that for different parent HTSCs,\ the nearest neighbor and
next--nearest neighbor exchange constants are similar, while the third
nearest neighbor $J$ appears to be material dependent. Furthermore, we find
that $J_{1}$ is insensitive to the hole doping, which agrees with recent
experiments and supports the implicit assumption of the t--J model with a
fixed magnitude of $J$. In contrast to the hopping integral, which strongly
depends on the position of the apical oxygen, our results show that the
magnetic exchange\ interaction is rather insensitive to the position of the
apical oxygen. In addtion to the LDA+U calculations, we studied the
dependence of the dynamic spin susceptibility $\chi (q,\omega )$ and
superconducting transition temperature T$_{c}$ on the exchange parameters in
a t--J model using a dynamic cluster approximation. We find that both $\chi
(q,\omega )$ and $T_{c}$ are only weakly affected by variations in the
exchange parameters beyond nearest neighbor. Based on these results, we
conclude that differences in long-range exchange terms between different
materials cannot explain their different superconducting transition
temperatures.

\section{ACKNOWLEDGMENT}

The authors acknowledge useful conversations with J. M. Dong, J. X. Li, G.
Kotliar, J. An and Q. H. Wang. X.G.W. acknowledges support from National Key
Project for Basic Researches of China (No. 2006CB921802), Natural Science
Foundation of China under Grant No. 10774067, and NSF of Jiangsu Province
through Grant No. BK2007127. We acknowledge the support from DOE SciDAC
Grant DE-FC0206ER25793. The work of S.Y.S. was supported by NSF DMR Grant
No. 0606498.


\begin{thebibliography}{99}
\bibitem{Pines} A.V. Chubukov and D. Pines, in \textit{The physics of
Superconductors} Vol. 2 (eds K.H. Bennemann and J.B. Ketterson) 495-590
(Springer, Berlin, 2003).

\bibitem{Monien} A.J. Millis, H. Monien, and D. Pines, Phys. Rev. B \textbf{%
42}, 167 (1990).

\bibitem{Anderson} P.W. Anderson, P.A. Lee, M. Randeria, T.M. Rice, N.
Trivedi, and F.C. Zhang, J. Phys.: Condens. Matt. \textbf{16} R755 (2004).

\bibitem{SO5} E. Demler, W. Hanke, and S.-C. Zhang, Rev. Mod. Phys. \textbf{%
76}, 909 (2004).

\bibitem{SFtheory} D. Scalapino, Phys. Rep. \textbf{250}, 329 (1995); T.
Moriya and K. Ueda, Rep. Prog. Phys. \textbf{66}, 1299 (2003).

\bibitem{Kanigel} R. Ofer, G. Bazalitsky, A. Kanigel, A. Keren, A. Auerbach,
J.S. Lord, and A. Amato, Phys. Rev. B \textbf{74}, 220508(R) (2006).


\bibitem{Dai} P. Dai, H.A. Mook, S.M. Hayden, G. Aeppli, T.G. Perring, R.D.
Hunt, F. Dogan, Science \textbf{284}, 1344 (1999).

\bibitem{LaCuO-1} K.B. Lyons, P.A. Fleury, J.P. Remeika, A.S. Cooper, and
T.J. Negran, Phys. Rev. B \textbf{37}, 2353 (1988).

\bibitem{LaCuO-2} S. M. Hayden, G. Aeppli, R. Osborn, A.D. Taylor, T.G.
Perring, S-W. Cheong, and Z. Fisk, Phys. Rev. Lett. \textbf{67}, 3622 (1991).

\bibitem{LCO-3} G. Aeppli \textit{et al}.,\ Phys. Rev. Lett. \textbf{62},
2052 (1989); R.R.P. Singh, P.A. Fleury, K.B. Lyons, and P.E. Sulewski, 
\textsl{ibid.} \textbf{62}, 2736 (1989); T. Thio \textit{et al.}, Phys. Rev.
B \textbf{38}, 905 (1988).

\bibitem{SW exp} R. Coldea\textit{\ et al.,\ }Phys. Rev. Lett. \textbf{86},
5377 (2001).

\bibitem{YBCO-1} J. M. Tranquada, G. Shirane, B. Keimer, S. Shamoto and M.
Sato, Phys. Rev. B \textbf{40}, 4503 (1989).

\bibitem{YBCO-2} S. Shamoto, M. Sato, J. M. Tranquada, B. J. Sternlieb, and
G. Shirane, Phys. Rev. B \textbf{48}, 13817 (1993).

\bibitem{Sr2CuOCl2} D. Vaknin, S.K. Sinha, C. Stassis, L.L. Miller, and D.C.
Johnston, Phys. Rev. B \textbf{41}, 1926 (1990).

\bibitem{Keimer} B. Keimer \textit{et al.}, Phys. Rev. B \textbf{45}, 7430
(1992).

\bibitem{hole-doped J} N.B. Christensen et al., Phys. Rev. Lett. \textbf{93}%
, 147002 (2004).

\bibitem{Ring-term exp} A. M. Toader\textit{, }J.P. Goff, M. Roger, N.
Shannon, J.R. Stewart, and M. Enderle,\textit{\ }Phys. Rev. Lett. \textbf{94}%
, 197202 (2005).

\bibitem{ring-term mix} J. Lorenzana, J. Eroles, and S. Sorella, Phys. Rev.
Lett. \textbf{83}, 5122 (1999).

\bibitem{ring term theory 1} I. de P. R. Moreira, C.J. Calzado, J.-P.
Malrieu, and F. Illas,\ Phys. Rev. Lett. \textbf{97}, 087003 (2006).

\bibitem{J.X. Li} T. Zhou, J.X. Li, and Z. D. Wang, Phys. Rev. B \textbf{75}%
, 054512 (2007); Y. Lubashevsky and A. Keren, \textsl{ibid. }\textbf{78},
020505 (2008).

\bibitem{Mag} S.M. Hayden \textit{et al}., Nature (London) \textbf{429}, 531
(2004); J.P. Hill \textit{et al.}, Phys. Rev. Lett. \textbf{100}, 097001
(2008).

\bibitem{J is important} E. Dagotto, Rev. Mod. Phys. \textbf{66}, 763 (1994).

\bibitem{J-Tc} M. Boninsegni and E. Manousakis, Phys. Rev. B \textbf{47},
11897 (1993); D.J. Scalapino and S.R. White, \textsl{ibid.} \textbf{58},
8222 (1998).

\bibitem{Renormalization} J. Igarashi, Phys. Rev. B \textbf{46}, 10763
(1992); R.R.P. Singh, \textsl{ibid.} \textbf{39}, 9760 (1989).

\bibitem{Theory LCO-1} R.L. Martin and F. Illas, Phys. Rev. Lett. \textbf{79}%
, 1539 (1997).

\bibitem{Theory LCO-2} D. Munoz, F. Illas, and I. de P.R. Moreira, Phys.
Rev. Lett. \textbf{84}, 1579\ (2000).

\bibitem{Theory LCO-3} J.F. Annett, R.M. Martin, A.K. McMahan and\ S.
Satpathy, Phys. Rev. B \textbf{40}, R2620 (1989).

\bibitem{Linear Spin-wave} P.W. Anderson, Phys. Rev. \textbf{86}, 694
(1952); R. Kubo, \textsl{ibid. }\textbf{87}, 568 (1952).

\bibitem{Shen} A. Damaselli, Z. Hussain, and Z.-X. Shen, Rev. Mod. Phys. 
\textbf{75}, 473 (2003).

\bibitem{t-t'-t''-U} J.-Y.P. Delannoy, M.J.P. Gingras, P.C.W. Holdsworth,
A.-M.S. Tremblay, cond-mat://0808.3167 (2008).

\bibitem{Qimiao Si} Q. Si, Y. Zha, K. Levin and J.P. Lu Phys. Rev. B \textbf{%
47}, 9055 (1993).

\bibitem{Wan J} X. Wan, Q. Yin and S.Y. Savrasov, Phys. Rev. Lett. \textbf{95%
}, 146602 (2006).

\bibitem{Savrasov-1996} S. Y. Savrasov, Phys. Rev. B \textbf{54} 16470
(1996).

\bibitem{LDA+U} V.I. Anisimov, J. Zaanen, and O.K. Andersen, Phys. Rev. B 
\textbf{44}, 943 (1991).

\bibitem{U J} V. I. Anisimov, M.A. Korotin, I.A. Nekrasov, Z.V. Pchelkina,
and S. Sorella, Phys. Rev. B \textbf{66}, 100502 (2002).

\bibitem{force theorem} A. I. Liechtenstein et al., J. Magn. Magn. Mater. 
\textbf{67}, 65 (1987); P. Bruno, Phys. Rev. Lett. \textbf{90}, 087205
(2003).

\bibitem{P-Tc} L. Gao \textit{et al.}, Phys. Rev. B \textbf{50}, 4260 (1994).

\bibitem{VCA} T. Thonhauser and C. Ambrosch-Draxl, Phys. Rev. B \textbf{67},
134508 (2003).

\bibitem{Apical O 1} J.A. Slezak et al., PNAS \textbf{105}, 3203 (2008).

\bibitem{Apical O 2} Y. Ohta, T. Tohyama and S. Maekawa, Phys. Rev. B 
\textbf{43}, 2968 (1991).

\bibitem{Andersen PRL} E. Pavarini, I. Dasgupta, T. Saha-Dasgupta, O.
Jepsen, and O.K. Andersen, Phys. Rev. Lett. \textbf{87}, 047003 (2001).

\bibitem{W.Ku} W.G. Yin and W. Ku, cond-mat://0702469.

\bibitem{hettler:dca} M.H.~Hettler, A.N. Tahvildar-Zadeh, M. Jarrell, T.
Pruschke, and H.R. Krishnamurthy, Phys. Rev B \textbf{58}, R7475 (1998);
M.H.~Hettler, M. Mukherjee, M. Jarrell, and H.R. Krishnamurthy, \textsl{ibid.%
} \textbf{61}, 12739 (2000).

\bibitem{ref:8} T.A.~Maier, M.~Jarrell, T.~Pruschke, and M.~Hettler, Rev.
Mod. Phys. \textbf{77}, 1027 (2005).

\bibitem{Kent} P.R.C. Kent, T. Saha-Dasgupta, O. Jepsen, O.K. Andersen, A.
Macridin, T.A. Maier, M. Jarrell, T.C. Schulthess, Phys. Rev. B \textbf{78},
035132 (2008).

\bibitem{jarrell_epl} M. Jarrell \textit{et al.}, Europhys. Lett. \textbf{56}%
, 563 (2001).

\bibitem{maier:pairmech} T.A. Maier, M.S. Jarrell, and D.J. Scalapino, Phys.
Rev. Lett. \textbf{96} 047005 (2006); Phys. Rev. B \textbf{74} 094513
(2006); \textsl{ibid.} \textbf{75}, 134519 (2007); T.A. Maier, A. Macridin,
M.S. Jarrell, and D.J. Scalapino, \textsl{ibid.} \textbf{76}, 144516 (2007).

\bibitem{maier:dcanca} Th. Maier \textit{et al.}, Eur. Phys. J B \textbf{13}%
, 613 (2000).

\bibitem{maier:edca} T.A. Maier, Physica B: Cond. Mat. \textbf{359-361},
512-514 (2005).

\bibitem{haule07} K. Haule and G. Kotliar, Phys. Rev. B \textbf{76}, 104509
(2007).

\bibitem{maier:pairglue} T.A. Maier, D. Poilblanc, and D.J. Scalapino, Phys.
Rev. Lett. \textbf{100}, 237001 (2008).
\end{thebibliography}
\end{document}